\documentclass[twocolumn,PRA,preprintnumbers,superscriptaddress,amsmath]{revtex4}
\usepackage[latin9]{inputenc}
\usepackage{textcomp}
\usepackage{amsmath}
\usepackage{amssymb}
\usepackage{graphicx}
\usepackage[unicode=true,pdfusetitle,
 bookmarks=true,bookmarksnumbered=false,bookmarksopen=false,
 breaklinks=false,pdfborder={0 0 1},backref=false,colorlinks=false]
 {hyperref}
\hypersetup{
 colorlinks,linkcolor=red,citecolor=blue}

\makeatletter
\@ifundefined{textcolor}{}
{%
 \definecolor{BLACK}{gray}{0}
 \definecolor{WHITE}{gray}{1}
 \definecolor{RED}{rgb}{1,0,0}
 \definecolor{GREEN}{rgb}{0,1,0}
 \definecolor{BLUE}{rgb}{0,0,1}
 \definecolor{CYAN}{cmyk}{1,0,0,0}
 \definecolor{MAGENTA}{cmyk}{0,1,0,0}
 \definecolor{YELLOW}{cmyk}{0,0,1,0}
}

\makeatother

\begin{document}

\title{Parametric down-conversion photon pair source on a nanophotonic chip}

\author{Xiang Guo}

\address{Department of Electrical Engineering, Yale University, New Haven,
Connecticut 06511, USA}

\author{Chang-Ling Zou}

\address{Department of Electrical Engineering, Yale University, New Haven,
Connecticut 06511, USA}

\author{Carsten Schuck}

\address{Department of Electrical Engineering, Yale University, New Haven,
Connecticut 06511, USA}

\author{Hojoong Jung}

\address{Department of Electrical Engineering, Yale University, New Haven,
Connecticut 06511, USA}

\author{Risheng Cheng}

\address{Department of Electrical Engineering, Yale University, New Haven,
Connecticut 06511, USA}

\author{Hong X. Tang}

\address{Department of Electrical Engineering, Yale University, New Haven,
Connecticut 06511, USA}
\begin{abstract}
\textbf{Quantum photonic chips, which integrate quantum light sources
alongside active and passive optical elements, as well as single photon
detectors, show great potential for photonic quantum information processing
and quantum technology. Mature semiconductor nanofabrication processes
allow for scaling such photonic integrated circuits to on-chip networks
of increasing complexity. Second order nonlinear materials are the
method of choice for generating photonic quantum states in the overwhelming
part of linear optic experiments using bulk components but integration
with waveguide circuitry on a nanophotonic chip proved to be challenging.
Here we demonstrate such an on-chip parametric down-conversion source
of photon pairs based on second order nonlinearity in an Aluminum
nitride microring resonator. We show the potential of our source for
quantum information processing by measuring high-visibility antibunching
of heralded single photons with nearly ideal state purity. Our down
conversion source operates with high brightness and low noise, yielding
pairs of correlated photons at MHz-rates with high coincidence-to-accidental
ratio. The generated photon pairs are spectrally far separated from
the pump field, providing good potential for realizing sufficient
on-chip filtering and monolithic integration of quantum light sources,
waveguide circuits and single photon detectors. }

\vbox{}\noindent\textbf{Keywords:} nanofabrication, quantum photonic
chip, second order nonlinear material, single photon source
\end{abstract}
\maketitle
\vbox{}

\noindent\textbf{INTRODUCTION}

\noindent Photons are excellent quantum information carriers because
they combine high-speed with long coherence times at room temperature
\cite{Kok2007_RMP,O'Brien2009_NO}. Photon pair sources based on spontaneous
parametric down conversion (SPDC) have enabled major advances in photonic
quantum computation and communication \cite{Burnham1970_PRL,Kwiat1995_PRL}.
Quantum teleportation \cite{Bouwmeester_1997,MaZeilinger_2012,BussieresGisin_2014},
multi-photon manipulation \cite{Peruzzo2010_Science,Matthews2009_NO,Shadbolt2011_NO,Huang2011_NC,Pan2012_RMP,Spring2013_Science,Broome2013_Science},
quantum algorithms \cite{Lu2007_PRL,Lanyon2007_PRL,Politi2009_Science,Martin-Lopez2012_NO,Zhou2013_NO}
and loophole-free tests of local realism \cite{GiustinaZeilinger_2015,ShalmNam_2015}
were all achieved with SPDC-photons generated in materials with a
strong second order ($\chi^{(2)}$) optical nonlinearity. Common nonlinear
crystals employed for SPDC, e.g. lithium niobate ($\mathrm{LiNbO_{3}}$),
are centimeter-sized and not yet compatible with standard nanofabrication
technologies \cite{Kruse2013_NJP,Solntsev2014_PRX,Jin2014_PRL}. Meanwhile,
silicon-based nanophotonic chips have emerged, which leverage mature
semiconductor fabrication processes to provide a stable and scalable
solution for optical quantum information processing \cite{O'Brien2009_NO,Tanzilli2012_LPR,Politi2008_Science,Pernice2012_NC,Schuck2015_NC}.
However, the generation of SPDC photons for current experiments with
silicon photonic chips is still realized in separate bulk optic setups.
One alternative approach resorts to exploiting the $\chi^{(3)}$ optical
nonlinearity of silicon, which allows for generating correlated photon
pairs via the process of spontaneous four wave mixing (SFWM) \cite{Sharping2006_OE,Clemmen2009_OE,Silverstone2014_NO}.
Integrating an optically pumped source of correlated photons with
waveguides and detectors on the same chip requires one, however, to
efficiently separate the generated signal and idler photons from the
co-propagating classical pump field. In the SFWM process the generated
photon pairs are spectrally separated by only a few nanometers from
the optical pump wavelength, which hence makes sufficient filtering
and integration of such sources with detectors on one chip a challenging
task. Because on-chip pump light suppression requires sophisticated
techniques \cite{Harris2014_PRX} a variety of important physical
processes could only be realized using SFWM in combination with off-chip
filtering, e.g. in source multiplexing \cite{Collins2013_NC}, on-chip
quantum interference \cite{Silverstone2014_NO}, and entanglement
\cite{Matsuda2012_SR,Grassani2015_Optica,Silverstone2015_NC}. SPDC
offers better prospects for efficient pump light suppression and consequently
integrating source and detector on the same chip. Efficient filtering
can for example be achieved by implementing a down conversion photon
pair source on a silicon substrate, where the pump light is in the
visible wavelength range and photon pairs are generated at infrared
wavelengths. The material absorption of silicon for visible light
is then large enough ($1740\,\mathrm{dB/cm}$) to guarantee efficient
suppression of undesired pump light before SPDC photons reach an on-chip
detector. 
\begin{figure*}[!]
\includegraphics{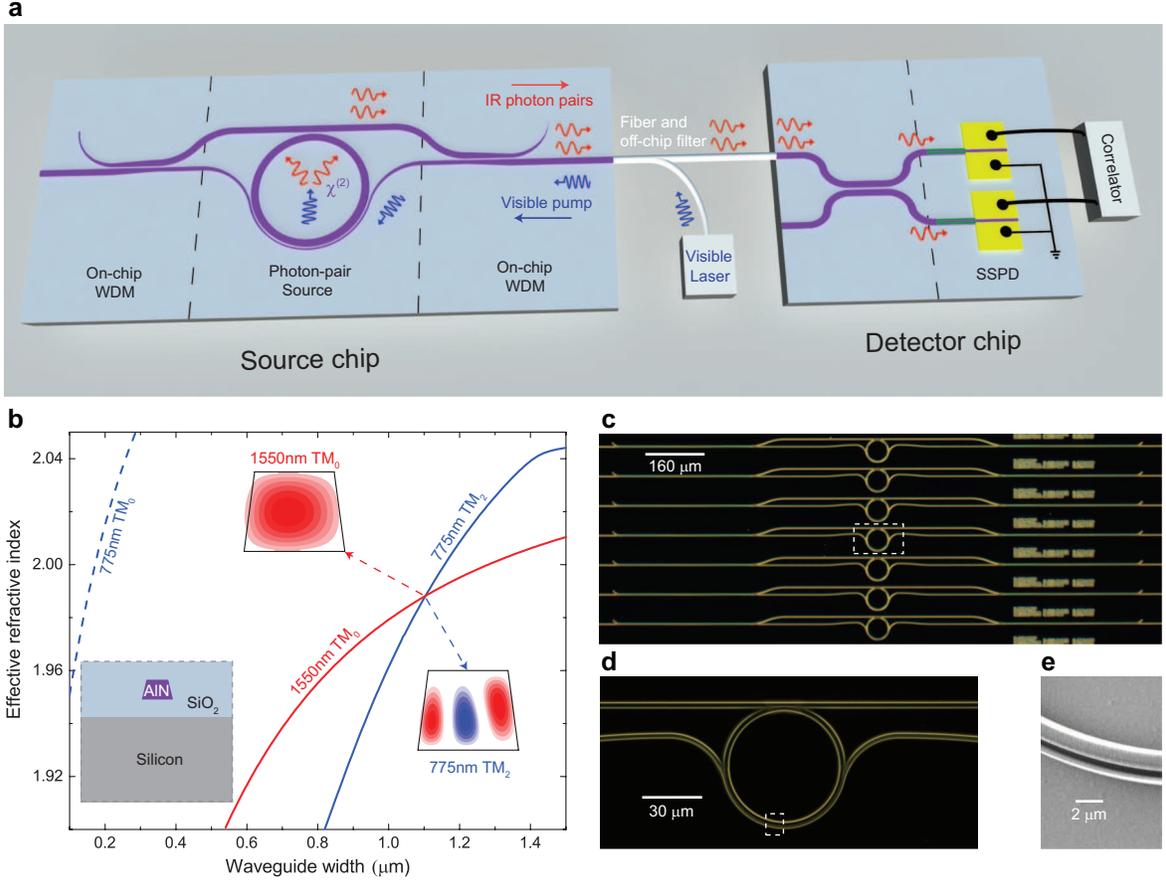}

\protect\caption{On-chip spontaneous parametric down conversion photon pair source.\textbf{
(a)} Schematic illustration of an on-chip photon pair source based
on a $\chi^{(2)}$ nonlinearity connected to superconducting single
photon detectors (SSPD) on another chip. Higher energy pump photons
(visible wavelengths) are coupled into a microring resonator and the
generated lower energy photon pairs (IR wavelengths) are randomly
split on the detector chip for coincidence measurements with integrated
SSPDs. A fiber coupled silicon filter (labeled as off-chip filter)
is used to reject the remaining pump light reflected back from fiber-to-chip
interface. \textbf{(b)} Effective refractive indices of modes in microring.
Phase match condition is satisfied with waveguide width around $1.10\,\mathrm{\mu m}$.
Lower left inset: the cross-section of the $\mathrm{AlN}$ waveguide.
Insets on the right: The electric field profiles at the cross-section
of microring of two phase-matched modes. \textbf{(c-e)} Device images
of $\mathrm{AlN}$ down conversion photon pair source. The white dashed
box shows the zoom-in region. \textbf{(c)} Optical image of an array
of microring photon pair source with on-chip WDMs. \textbf{(d)} Zoom-in
of a single microring resonator with both IR (top) and visible (bottom)
bus waveguides. \textbf{(e)} SEM picture showing the coupling region
of visible light bus waveguide (narrow) with microring resonator (wide).}
\label{Fig1}
\end{figure*}

Aluminum nitride ($\mathrm{AlN}$) is a new material suitable for
scalable photonic integrated circuits \cite{Xiong2012_NJP,Pernice2012_APL}.
Its strong intrinsic second order nonlinearity ($\chi^{(2)}$) not
only shows great potential for realizing on-chip photon pair sources
based on SPDC \cite{Pernice2012_APL}, but also permits integrated
low-loss, high-speed electro-optic (EO) phase modulation \cite{Xiong2012_NJP}.
Here we demonstrate an integrated SPDC photon pair source using a
high quality factor ($Q$) $\mathrm{AlN}$ microring resonator. For
compatibility with telecommunication technology we realize nonlinear
conversion of visible wavelength ($775\,\mathrm{nm}$) pump photons
to telecom wavelength ($1550\,\mathrm{nm}$) photon pairs. High refractive
index contrast between the AlN-waveguides and a silicon dioxide ($\mathrm{SiO_{2}}$)
cladding layer allows for small device footprint and enables dense
integration on silicon handles. The generated photon pairs are characterized
by waveguide coupled superconducting single photon detectors (SSPD)
integrated on a dedicated chip. We achieve high photon pair production
rates of $3\,\mathrm{MHz/mW}(5.8\,\mathrm{MHz/mW})$ for degenerate
(non-degenerate) down conversion. The overall photon pair emission
rate is more than $20\,\mathrm{MHz/mW}$, which is comparable to the
state-of-the-art bulk optic photon pair sources \cite{Jin2014_PRL}.
In terms of the spectral brightness, the demonstrated $\mathrm{AlN}$
microring SPDC source is much brighter than conventional waveguide
or bulk crystal based sources. We observe high-visibility antibunching
of heralded single photons, characterized by second order intensity
correlations of $g_{h}^{(2)}(0)=0.088\text{\textpm}0.004$. The suitability
of our SPDC source for quantum information applications is further
highlighted by nearly ideal purity of the heralded photons, which
we demonstrate in self-correlation measurements of the idler photons.

\vbox{}

\vbox{}

\noindent\textbf{MATERIALS AND METHODS}

\noindent\textbf{Experimental setup and device engineering}

\noindent The device is shown in Fig.$\,$\ref{Fig1}a. We realize
the SPDC source as a high quality factor $\mathrm{AlN}$ microring
resonator, which enhances the pump photon interaction with the material\textquoteright s
$\chi^{(2)}$ nonlinearity \cite{Yang2007_OL}. This allows for producing
down-conversion photon pairs with long-coherence time at low optical
pump power. Here, the visible wavelength pump laser field is guided
into the microring resonator via a narrow wrap-around waveguide, while
the generated IR photon pairs are coupled out via a wider bus waveguide.
We design an on-chip wavelength division multiplexer (WDM), which
guides IR photons back into the optical fiber towards the detectors.
After passing through a fiber coupled silicon filter, the residual
visible pump photons are rejected while IR photons are guided to waveguide
coupled superconducting single photon detectors (SSPD), which are
integrated on a separate chip \cite{Pernice2012_NC,Schuck2015_NC,Schuck2013_SR}
inside a cryostat. Waveguide directional couplers on this detection
chip allow for 50/50 splitting of photon pairs before detection with
the SSPDs and signal analysis with time correlated single photon counting
(TCSPC) electronics.

The lower left inset in Fig.$\,$\ref{Fig1}b shows the cross-section
of the $\mathrm{AlN}$ chip. $\mathrm{AlN}$ forms the core of the
waveguide while $\mathrm{SiO_{2}}$ acts as a low refractive index
cladding layer, on top of the silicon substrate. A degenerate (non-degenerate)
SPDC process involves one optical pump mode in the visible wavelength
band and one (two) signal and idler modes in the IR wavelength band.
Energy conservation implies the condition $\omega_{\mathrm{vis}}=\omega_{\mathrm{IR,1}}+\omega_{\mathrm{IR,2}}$,
while momentum conservation requires $m_{\mathrm{vis}}=m_{\mathrm{IR,1}}+m_{\mathrm{IR,2}}$,
where $\omega_{x}$ and $m_{x}$ ($x=\mathrm{vis,\,IR,1\,or\,IR,2}$)
are the frequencies and azimuthal numbers of the visible and IR modes,
respectively (for degenerate SPDC $\omega_{\mathrm{IR,1}}=\omega_{\mathrm{IR,2}}$
and $m_{\mathrm{IR,1}}=m_{\mathrm{IR,2}}$). To fulfill these two
conditions and realize efficient nonlinear conversion, it is necessary
to match the effective refractive indices $n_{eff}=m_{x}c/\omega_{x}r$
($c$ is the speed of light in vacuum and $r$ is the radius of the
microring) of the visible pump and IR signal and idler modes. This
phase matching condition can be satisfied for a higher order transverse
magnetic (TM) visible wavelength pump mode and fundamental TM ($\mathrm{TM_{0}}$)
signal and idler modes in the IR \cite{Pernice2012_APL,Levy2011_OE}.
In Fig.$\,$\ref{Fig1}b we show how this effective refractive index
matching is achieved for a $775\,\mathrm{nm}$ $\mathrm{TM_{2}}$
mode and a $1550\,\mathrm{nm}$ $\mathrm{TM_{0}}$ mode by engineering
the waveguide-width of the $\mathrm{AlN}$ microring. In the right
two insets of Fig.$\,$\ref{Fig1}b the corresponding mode profiles
in a $1.10\,\mu m$ width waveguide are shown. 

The high refractive index contrast between waveguide ($\mathrm{AlN}$)
and cladding ($\mathrm{SiO_{2}}$) materials allows for a small device
footprint and dense integration. In Fig.$\,$\ref{Fig1}c we show
an optical micrograph of an array of $\mathrm{AlN}$ microring photon
pair sources integrated with on-chip WDMs. Two independent waveguide
channels are designed for visible and IR light, respectively (Fig.$\,$\ref{Fig1}d).
To excite the visible wavelength $\mathrm{TM_{2}}$ mode inside the
ring resonator we utilize a narrow wrap-around waveguide. Adjusting
the gap between the ring and the wrap-around waveguide as well as
the wrap-around waveguide width it is possible to adiabatically couple
the $\mathrm{TM_{0}}$ pump mode of the feed waveguide to the $\mathrm{TM_{2}}$
mode of the microring under critical coupling conditions. As long
as the gap between the wrap-around waveguide and the microring is
large enough, the existence of this narrow waveguide will not deteriorate
the quality factor for the ring resonator\textquoteright s IR modes
(see supplementary section I). The wrap-around waveguide is tapered
down to $100\,\mathrm{nm}$, as shown in the SEM image of the coupling
region (Fig.$\,$\ref{Fig1}e) while the coupling gap between wrap-around
waveguide and microring is $500\,\mathrm{nm}$.

\vbox{}

\noindent\textbf{Device fabrication and measurement}

\noindent A $\mathrm{1\,\mu m}$ thin $\mathrm{AlN}$ film is sputtered
on a commercial oxide-on-silicon wafer. We use FOX-16 resist and define
patterns in electron beam lithography. After development in MF312
developer, $\mathrm{Cl_{2}/BCl_{3}/Ar}$ chemistry is used to etch
into the $\mathrm{AlN}$ layer. The chip is then coated with a $\mathrm{2.5\,\mu m}$
of PECVD oxide cladding layer as protection during subsequent polishing
steps. The device is finally annealed in an $\mathrm{O}_{2}$ atmosphere
for $\mathrm{5\,\mathrm{h}}$ at $\mathrm{950\,\textdegree C}$ to
improve the quality of PECVD oxide. The detector chip is fabricated
from a commercial $330\,\mathrm{nm}$ $\mathrm{SiN}$-on-insulator
wafer onto which we sputter an $8\,\mathrm{nm}$ thin film of $\mathrm{NbTiN}$.
Electrode pads are defined in PMMA resist via electron beam lithography
followed by gold deposition and lift-off in acetone. In a second electron
beam lithography step the SSPD nanowires are patterned in HSQ negative
tone resist and transferred into the $\mathrm{NbTiN}$ layer using
$\mathrm{CF_{4}}$ chemistry. In the third and final electron beam
lithography step the $\mathrm{SiN}$ waveguides are written in ZEP
positive tone resist followed by a timed reactive ion etch in $\mathrm{CHF_{3}/O_{2}}$
chemistry.

We use a continuous-wave visible wavelength laser (TLB-6712) to pump
the SPDC source. A fiber coupled silicon absorber (OZ optics) is used
to filter out the pump light reflected from fiber-to-chip interface.
We achieve $80\,\mathrm{dB}$ attenuation for pump light at $3\,\mathrm{dB}$
insertion loss for IR light. For the degenerate SPDC coincidence measurement,
a band-pass tunable filter is used to spectrally filter out the degenerate
down conversion photons. For the measurements of the SPDC thermal
state, a dense wavelength division multiplexer (DWDM) is used to select
the idler branch from the nearest non-degenerate down conversion photon
pair. The photons are then sent into the detector chip, where superconducting
single photon detectors are integrated with a 50/50 directional coupler
for self-correlation measurements \cite{Schuck2015_NC}. Electrical
signals from the on-chip detectors are sent into a Time-Correlated
Single Photon Counting (TCSPC) system (Picoharp 300) for time tagging
\cite{Schuck2013_APL}. For non-degenerate cross-correlation measurements,
a DWDM is used to separate signal and idler photons, which are then
sent to two separate on-chip detectors for coincidence measurement.
\begin{figure}[tp]
\includegraphics[width=8.8cm]{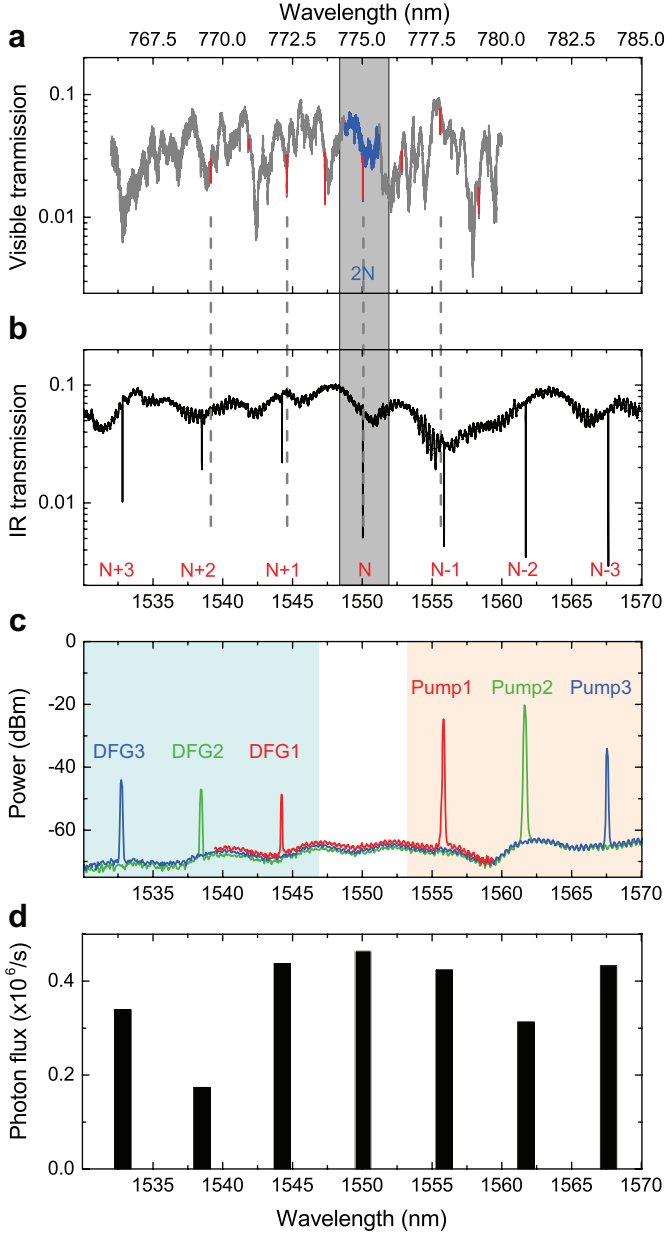} \protect\caption{Linear and nonlinear characteristics of the $\chi^{(2)}$ source.
\textbf{(a-b)} Transmission spectra of the microring resonator. Four
gray dashed lines are aligned with the visible $\mathrm{TM_{2}}$
mode resonances. Gray region shows the phase-matched mode pair ($\mathrm{TM_{2,2N}}$
and $\mathrm{TM_{0,N}}$) for SHG and degenerate SPDC, whose resonances
are aligned within the linewidth of IR mode resonance. \textbf{(a)}
Visible light transmission spectrum, with $\mathrm{TM_{2}}$ mode
resonances emphasized by red lines. \textbf{(b)} IR light transmission
spectrum, with $\mathrm{TM_{0}}$ modes identified by the azimuthal
mode number. \textbf{(c)} Difference frequency generation (DFG) measured
by optical spectrum analyzer. Peaks due to input IR pump lasers are
shown in the light orange region and generated DFG signals are shown
in the light green region. Visible pump laser with $1.9\,\mathrm{mW}$
power on-chip is fixed on resonance with $\mathrm{TM_{2,2N}}$ mode.
\textbf{(d)} Single photon flux arriving at the detector chip from
the degenerate and nearest three groups of non-degenerate down conversion.
The counts are calibrated by the wavelength dependent detection efficiency
(Supplementary Section II).}
\label{Fig2}
\end{figure}

\vbox{}

\noindent\textbf{RESULTS AND DISCUSSION}

\noindent\textbf{Characterization of on-chip down-conversion source}

\noindent We first characterize the classical performances of the
microring resonator. Figure$\,$\ref{Fig2}a shows the microring transmission
spectrum for visible light in a slightly under-coupled configuration
(a critically-coupled spectrum is shown in the supplementary section
I). Figure$\,$\ref{Fig2}b shows the transmission spectrum for IR
light, when the bus waveguide is critically-coupled to the microring
resonator. The two spectra show how the $\mathrm{TM_{2,2N}}$ mode
at $775\,\mathrm{nm}$ aligns with the $\mathrm{TM_{0,N}}$ mode at
twice the wavelength according to energy conservation ($\omega_{vis}=2\omega_{IR}$).
Here $\mathrm{N(2N)}$ stands for the azimuthal mode number, which
is determined by momentum conservation ($m_{\mathrm{vis}}=2m_{\mathrm{IR}}$)
in the degenerate SPDC process. To perfectly fulfill the energy conservation
($\omega_{vis}=2\omega_{IR}$), we vary the global temperature of
the chip and determine the optimized working temperature using second
harmonic generation (SHG) \cite{Xiong2012_NJP,Pernice2012_APL} as
a figure of merit, which is directly related to SPDC efficiency (see
theoretical derivation in supplementary section III). By pumping at
$\mathrm{TM_{0,N}}$ mode and monitoring the output SHG from $\mathrm{TM_{2,2N}}$
mode, we observe a maximum SHG efficiency of $\eta_{SHG}=P_{\mathrm{SHG}}/P_{\mathrm{p}}^{2}=1.16\,\mathrm{W}^{-1}$,
where $P_{\mathrm{SHG}}$($P_{p}$) is the optical power of the generated
second harmonic light (pump laser). Due to the group velocity mismatch
between visible and IR modes, energy conservation cannot be satisfied
for SHG in other modes, as can be seen in Figs.$\,$3a and b, where
no pairs of IR and visible light modes are aligned except for the
$\mathrm{TM_{2,2N}}$ and $\mathrm{TM_{0,N}}$ modes. However, the
group velocity dispersion (GVD) is relatively small in the IR region,
and energy conservation can be fulfilled for IR modes over a wide
wavelength range for difference/sum frequency generation (DFG/SFG)
and conversely for non-degenerate SPDC. We verify this by fixing the
visible wavelength pump laser at the resonance of the $\mathrm{TM_{2,2N}}$
mode and sweep the IR probe laser over a wavelength range covering
the neighboring resonances (e.g. $\mathrm{TM_{0,N-1}/TM_{0,N-2}/TM_{0,N-3}}$
modes). We observe DFG of various modes, as shown in Fig.$\,$\ref{Fig2}c.
The fact that both SHG and various DFG-configurations can be observed
indicates the possibility of generating both degenerate and non-degenerate
photon-pairs in wavelength bands spaced similar to a frequency-comb. 

We use SSPDs to characterize the statistical properties of photons
generated by the SPDC source. In Fig.$\,$\ref{Fig2}d the photon
flux for different ring resonances is shown for fixed power and wavelength
of the visible light pump laser. The variation in the detected photon
rates for different resonance wavelengths can be explained by the
difference in the quality factors of the IR resonator modes because
the photon-pair generation rate is linearly dependent on the quality
factor. As discussed in the supplementary information, the bandwidth
for non-degenerate SPDC can be as large as $40\,\mathrm{nm}$. 
\begin{figure}[tph]
\includegraphics[width=8.8cm]{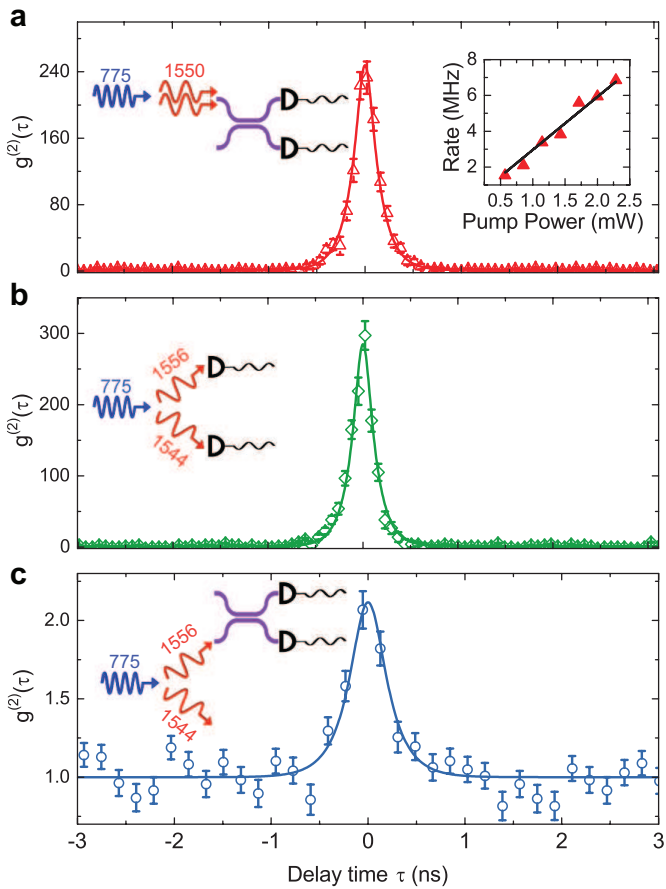} \protect\caption{Second order correlation function ($g^{(2)}(\tau)$) of down converted
photons. \textbf{(a)} Self-correlation measurement of degenerate down
conversion photon pairs. Right inset: Power dependence of degenerate
photon pair generation rate. Normalized Generation rate of $3.0\,\mathrm{MHz/mW}$
has been fitted. \textbf{(b)} Cross-correlation measurement of the
nearest non-degenerate down conversion photon pairs. \textbf{(c)}
Self-correlation measurement of the idler in the nearest non-degenerate
down conversion. The left insets in \textbf{(a-c)} show the schematics
of each measurement. The solid lines are fittings with the convolution
function between a double exponential decay and a Gaussian distribution
function related to the chosen bin-width and detector jitter.}
\label{Fig3}
\end{figure}

\vbox{}

\noindent\textbf{Correlated photon-pair generation}

\noindent The statistical properties of the generated photon-pairs
can be analyzed in terms of the second order correlation function
(\textbf{$g^{(2)}$}), which we measure using two detectors at the
outputs of a beam splitter and normalize to photon flux \cite{Fortsch2015_PRA}
\begin{equation}
g^{(2)}(\tau)=\frac{R_{cc}(\tau)}{R_{1}R_{2}\tau_{b}}
\end{equation}
where $R_{cc}$ is the coincidence count rate, $R_{1}$ ($R_{2}$)
is the count rate of detector 1(2), and $\tau_{b}$ is the coincidence
time window. Here the coincidence rate $R_{cc}$ is a function of
photon-photon arrival time delay $\tau$, while $R_{1}$, $R_{2}$
and $\tau_{b}$ are constant, which combine to the accidental coincidence
rate $R_{ac}=R_{1}R_{2}\tau_{b}$ . We note that the $g^{(2)}(\tau)$
function is a rescale of coincidence rate $R_{cc}(\tau)$ by accidental
coincidence rate $R_{ac}$. While the value of $R_{cc}(\tau)$ and
$R_{ac}$ are both dependent on the losses of the measuring systems,
the value of $g^{(2)}(\tau)$ is independent of system losses, allowing
us to extract the pair generation rate and SPDC photon bandwidth from
the measured $g^{(2)}(\tau)$ function directly.

For degenerate SPDC photon pairs we measure the self-correlation function
with two waveguide coupled SSPDs integrated with a 50/50 directional
coupler, as sketched in the left inset of Fig.$\,$\ref{Fig3}a. We
observe a clear coincidence peak centered at zero delay time, which
indicates strong temporal correlations between the photons emitted
from the SPDC source. The second order self-correlation function for
degenerate SPDC photon-pairs is given by \cite{Fortsch2013_NC,Reimer2014_OE,Clausen2014_NJP}
\begin{equation}
g_{self}^{(2)}(\tau)=1+\frac{1}{4R\tau_{\mathrm{c}}}e^{-\left|\tau\right|/\tau_{\mathrm{c}}},\label{eq:1-1}
\end{equation}
where $\tau$ is the delay time between two photons, $\tau_{c}$ is
the coherence time of the SPDC photons, and $R$ is the photon-pair
generation rate. Here, random fluctuations in the photons arrival
time ($\delta\tau$) have to be taken into account because the detector
jitter ($\tau_{j}$, see supplementary section V) and the coincidence
time window ($\tau_{b}$) are not negligible compared with the photons\textquoteright{}
coherence time ($\tau_{c}$). Assuming the arrival time fluctuations
$\delta\tau$ follow a normal distribution $\frac{1}{\sqrt{2\pi}\tau_{\mathrm{w}}}e^{-\delta\tau^{2}/2\tau_{\mathrm{w}}^{2}}$,
with standard deviation $\tau_{\mathrm{w}}=\sqrt{2\tau_{\mathrm{j}}^{2}+(\frac{\tau_{\mathrm{b}}}{2})^{2}}$,
the correlation function can be expressed as (see supplementary section
V)
\begin{align}
g_{self}^{(2)}(\tau) & =1+\frac{1}{8R\tau_{\mathrm{c}}}e^{\tau_{w}^{2}/2\tau_{\mathrm{c}}^{2}}\left[f_{+}(\tau)+f_{-}(\tau)\right],
\end{align}
where $f_{\pm}(\tau)=\left[1\mp\mathrm{erf}\left(\frac{\tau\pm\tau_{\mathrm{w}}^{2}/\tau_{\mathrm{c}}}{\sqrt{2}\tau_{\mathrm{w}}}\right)\right]\cdot e^{\pm\tau/\tau_{\mathrm{c}}}$
and $\mathrm{erf}(x)$ is the error function. From a fit to the data
in Fig.$\,$\ref{Fig3}a, we obtain the degenerate photon pair generation
rate $R=5.9\,\mathrm{MHz}$ for $1.9\,\mathrm{mW}$ pump power on-chip.
The bandwidth of the photons extracted from the fit to the data is
$\Delta\nu=\frac{1}{2\pi\tau_{c}}=1.1\,\mathrm{GHz}$, which agrees
with the measured linewidth of the IR resonator mode. The photon-pair
generation rate as a function of pump power is shown in the right
inset of Fig.$\,$\ref{Fig3}a. We obtain a generation rate of $3.0\,\mathrm{MHz/mW}$
for degenerate SPDC.

For non-degenerate SPDC photon pairs, the photons of a pair may have
different wavelengths. If we discard all the signal photons and only
measure the emission of idler photons, a thermal state statistics
is expected \cite{Blauensteiner2009_PRA}. We measure the self-correlation
function for the nearest idler ($\mathrm{TM_{0,N-1}}$) photons, as
shown in the inset of Fig.$\,$\ref{Fig3}c. We obtain $g^{2}(0)=2.07\pm0.12$
from a fit to the data in Fig.$\,$\ref{Fig3}c, which is in agreement
with the expected value of $g^{2}(0)=2$ for single mode thermal state
\cite{Fortsch2015_PRA,Christ2011_NJP}. 

We then separate signal and idler photons by a DWDM and use two independent
waveguide coupled SSPDs to measure the second order cross-correlation
function between signal and idler, as shown in the inset of Fig.$\,$\ref{Fig3}b.
For non-degenerate photon pairs, the cross-correlation function is
given as: 
\begin{align}
g_{cross}^{(2)}(\tau) & =1+\frac{1}{4R\tau_{\mathrm{c}}}e^{\tau_{w}^{2}/2\tau_{\mathrm{c}}^{2}}\left[f_{+}(\tau)+f_{-}(\tau)\right].
\end{align}
From a fit to the data in Fig.$\,$\ref{Fig3}b we extract the photon
pair generation rate of $R=11.0\,\mathrm{MHz}$ for $1.9\,\mathrm{mW}$
pump power ($5.8\,\mathrm{MHz/mW}$) and the bandwidth $\Delta\nu=\frac{1}{2\pi\tau_{c}}=1.1\,\mathrm{GHz}$,
which is similar to the degenerate SPDC case. We conclude that non-degenerate
SPDC is around two times more efficient than degenerate SPDC, which
matches well with our theoretical calculation and originates from
different coefficients of interaction Hamiltonian for degenerate and
non-degenerate SPDC processes (supplementary section III). Additional
cross-correlation measurements for photon pairs emitted into other
microring resonance modes are shown in the supplementary information.
The total (including degenerate and different groups of non-degenerate)
photon pair generation rate for this $\mathrm{AlN}$ microring source
is more than $20\,\mathrm{MHz/mW}$, which is comparable or even higher
than the state-of-the-art SPDC photon pair source using bulk or waveguide-based
$\chi^{(2)}$ crystals \cite{Jin2014_PRL}. In terms of the spectral
brightness, $\mathrm{AlN}$ microring source is much brighter due
to microring's narrow linewidth. 
\begin{figure}[tph]
\includegraphics[width=8.8cm]{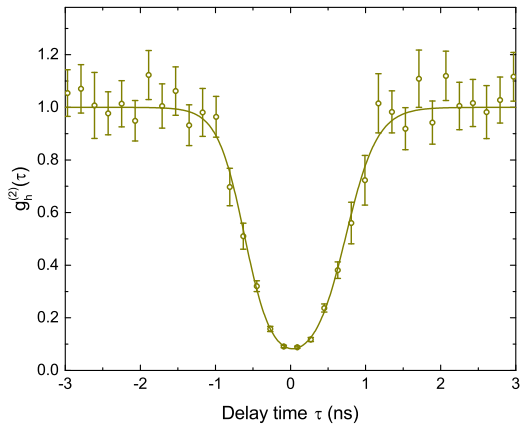} \protect\caption{Heralded signal-signal correlation function ($g_{h}^{(2)}(\tau)$)
of down converted photons. On-chip pump power of $1.9\,\mathrm{mW}$
is used to get enough counting rate. Here the heralded anti-bunching
dip of $g_{h}^{(2)}(0)=0.088\pm0.004$ indicates that the photon pair
source is operating in the single photon regime.}
\label{Fig4}
\end{figure}

\vbox{}

\noindent\textbf{Source of heralded single photons}

\noindent With $1.9\,\mathrm{mW}$ pump power on-chip, the photon
pair generation rate for the nearest non-degenerate modes ($\mathrm{TM_{0,N+1}}$
and $\mathrm{TM_{0,N-1}}$) is $11.0\,\mathrm{MHz}$. To further verify
that our photon pair source can be used as a heralded single photon
source, we measure the normalized signal-signal self-correlation function
conditioned on the detection of an idler photon \cite{Fortsch2013_NC,Bettelli2010_PRA}
\begin{equation}
g_{h}^{(2)}(t_{s1},t_{s2}|t_{i})=\frac{P_{ssi}(t_{s1},t_{s2},t_{i})}{r(0)^{3}g_{si}^{(2)}(t_{s1}-t_{i})\cdot g_{si}^{(2)}(t_{s2}-t_{i})},
\end{equation}
where $P_{ssi}$ is the coincidence rate of detecting one idler and
two signal photons, $g_{si}^{(2)}$ is the second order cross correlation
function and $r(\tau)$ is first order correlation function \cite{Bettelli2010_PRA}.
Here we are interested in the special case of $g_{h}^{(2)}(\tau)\equiv g_{h}^{(2)}(0,\tau|0)$.
For an ideal photon pair source, we expect to detect at most one signal
photon upon detection of one (heralding) idler photon. Thus an antibunching
dip around zero delay time ($\tau=0$) is expected. In our experiment,
we split up the nearest non-degenerate photon pairs deterministically
using a DWDM. We then use the idler photon as a herald for the detection
of its partner photon and measure the autocorrelation function for
the latter. Shown in Fig.$\,$\ref{Fig4} is the measurement result
with no background noises or dark counts subtracted. We extract $g_{h}^{(2)}(0)=$
$0.088\pm0.004\ll0.5$ from the data, which confirms that our SPDC
source indeed yields single photons with nonclassical correlations. 

\vbox{}

\noindent\textbf{Device losses and measuring efficiencies}

\noindent The raw data of the coincidence measurements is shown in
the supplementary information. The measured coincidence rate is around
$80\,\mathrm{Hz}$. For a pump power of $1.9\,\mathrm{mW}$ on-chip
we infer an emission rate of photon pairs in the nearest non-degenerate
modes of $11.0\,\mathrm{MHz}$ from a fit to the coincidence data
in Fig.$\,$\ref{Fig3}b. The difference between these rates arises
from losses at the various device interfaces, i.e. microring-to-waveguide
($3\,\mathrm{dB}$), fiber-to-chip interface ($3.5\,\mathrm{dB}$),
silicon filter ($3\,\mathrm{dB}$), DWDM ($6\,\mathrm{dB}$) and non-ideal
detector efficiency ($10\,\mathrm{dB}$). Without subtracting any
background noise or dark counts we find a coincidence-to-accidental
ratio (CAR) of $560$ for a minimum pump power of $0.6\,\mathrm{mW}$,
at which the on-chip generation rate amounts to $3.5\,\mathrm{MHz}$. 

\vbox{}

\noindent\textbf{On-chip wavelength division multiplexer}

\begin{figure}[tp]
\includegraphics[width=8.8cm]{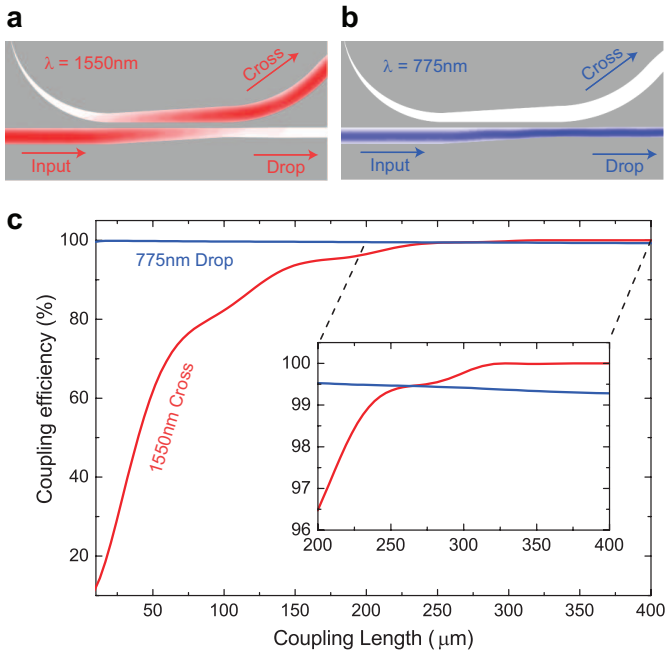} \protect\caption{On-chip wavelength division multiplexer (WDM). \textbf{(a-b)} Simulated
electric field profiles of the designed on-chip WDM device for IR
($\lambda=1550\,\mathrm{nm}$) and visible ($\lambda=775\,\mathrm{nm}$)
light. \textbf{(c)} Coupling efficiency of IR light to cross port
and visible light to drop port. Inset, Zoom in of coupling efficiency
for coupling length ranging from $200\,\mathrm{\mu m}$ to $400\,\mathrm{\mu m}$.}
\label{Fig5}
\end{figure}

\noindent The large difference in pump and signal / idler wavelengths
in the SPDC process allows for a variety of design choices to separate
pump light from the generated photons. One could choose to exploit
wavelength selective material absorption, e.g. silicon will strongly
absorb visible light but transmit IR light. Another approach is to
design WDM waveguide circuits. In this work we choose to realize the
latter one because a WDM structure has the additional benefit of simplifying
the fiber-to-chip coupling interface: 1) When characterizing the IR
and visible performances of the microring, only one optical fiber
at each side of the photonic chip is needed. 2) When characterizing
the down-conversion photon pairs, a single optical fiber is simultaneously
used to send the pump light into and collect generated photon pairs
from the chip (as shown in Fig.$\,$\ref{Fig1}a). The designed WDM
structure employs tapered waveguide couplers, as shown in Figs.$\,$\ref{Fig5}a
and b. In dielectric waveguides the optical mode confinement decreases
towards longer wavelengths, such that IR light will have a longer-range
evanescent field outside the waveguides as compared to visible light.
For two waveguides in close proximity the resulting coupling is hence
stronger for IR than for visible wavelength light. We hence adjust
the coupling length such that IR light is efficiently transferred
from one waveguide to another (Fig.$\,$\ref{Fig5}a) while visible
light remains unaffected and is transmitted through the coupling region
without coupling to the neighboring waveguide (Fig.$\,$\ref{Fig5}b). 

Here the adiabatic taper WDM design is realized with large fabrication
tolerances and broadband working wavelength. The relationship between
coupling efficiency and coupling length is shown in Fig.$\,$\ref{Fig5}c
for $\lambda=1550\,\mathrm{nm}$ and $\lambda=775\,\mathrm{nm}$.
The design shown in Figs.$\,$\ref{Fig5}a and b corresponds to a
coupling length of $350\,\mathrm{\mu m}$ and a gap of $0.4\,\mathrm{\mu m}$.
With increasing coupling length, the cross port coupling for IR light
increases monotonically and saturates at 100\%, while the visible
light transmission into the drop port decreases linearly with coupling
length but remains above 99\% over the entire range. The inset of
Fig.$\,$\ref{Fig5}c shows that more than 99\% coupling efficiency
into the cross and drop ports are achieved simultaneously for IR and
visible light, respectively, for coupling lengths ranging from $250\,\mathrm{\mu m}$
to $400\,\mathrm{\mu m}$. This performance corresponds to $20\,\mathrm{dB}$
suppression of visible pump light with less than $0.044\,\mathrm{dB}$
insertion loss for IR light for each of the on-chip WDM structures.
In future implementations, a cascade of such WDMs could be used for
realizing sufficient suppression of guided pump light. Note that in
the current experimental configuration, a portion of the pump light
is directly reflected back into the optical fiber from the fiber-to-chip
interface and is thus guided towards the detectors. To suppress these
residual pump photons we use an additional fiber-coupled silicon filter,
i.e. off-chip.

\vbox{}

\noindent\textbf{Discussion}

\noindent The high-visibility antibunching of heralded single photons
shown in Fig. \ref{Fig4} confirms the nonclassical character of our
SPDC source. However, for quantum information processing it is furthermore
desirable to generate photons in a pure state \cite{Spring2013_OE}.
We use the Schmidt number, $K$, to describe any remaining entanglement
between optical modes of a SPDC pair. Schmidt number $K$, heralded
single photon state purity $P$ and the second order self-correlations
are related via $g^{(2)}(0)=1+1/K=1+P$ \cite{Christ2011_NJP,Spring2013_OE}.
Ideal heralded purity ($P=1$) is obtained if all measured photons
are found in the same optical mode ($K=1$). Conversely, if photons
are found in multiple output modes ($K\gg1$), the correlation function
$g^{(2)}(0)$ approaches 1 and the heralded purity approaches 0. In
our experimental configuration, down-converted photons will only be
emitted into $\mathrm{TM_{0}}$ modes because other spatial modes
do not fulfill the phase matching condition. These $\mathrm{TM_{0}}$
modes are spectrally separated by at least one free spectral range
(FSR) , as shown in Fig.$\,$\ref{Fig2}d. Selecting one of these
modes with a bandpass filter thus yields single mode emission. The
measured $g^{2}(0)=2.07\pm0.12$ ($P=1.07\pm0.12$) in Fig.$\,$\ref{Fig3}c
confirms that single mode emission and nearly ideal (unit) heralded
purity are indeed achieved by our SPDC source.

In addition to the purity of generated photons, sufficient on-chip
pump rejection is a key requirement for achieving source-detector
integration on the same chip. Here we discuss the prospects of realizing
sufficient on-chip filtering. 1) Suppressing guided pump light in
photonic waveguide circuits: The filtering of pump photons propagating
inside a waveguide could for example be achieved by depositing a thin
layer of silicon on top of the waveguide interfacing source (circuit)
and detector. Silicon has an absorption coefficient of $1740\,\mathrm{dB/cm}$
for the pump wavelength ($775\,\mathrm{nm}$) at cryogenic temperatures,
while being transparent for down-converted photons (around $1550\,\mathrm{nm}$).
The hybrid $\mathrm{AlN}$-silicon waveguide hence results in great
selective absorption of pump photons. Alternatively, five cascaded
on-chip WDMs (as described above) could also provide efficient pump
light rejection. 2) Suppressing unguided pump photons propagating
in free-space, substrate and cladding materials: Photons incident
from free space can efficiently be absorbed in a metal layer covering
the area where detectors are located \cite{Reithmaier2015_NL}. Pump
photons scattered into a transparent substrate on the other hand pose
a significant challenge in SFWM experiments, where the pump light
in telecom band near $1550\,\mathrm{nm}$ can propagate losslessly
in the cladding (usually $\mathrm{SiO_{2}}$) and substrate (usually
silicon) materials and finally couple to the detection region, limiting
the on-chip filter\textquoteright s performance \cite{Harris2014_PRX}.
In our case, however, the visible pump photons scattered into the
silicon substrate layer are efficiently absorbed due to large material
absorption. Pump photons coupled directly from the input fiber into
$\mathrm{SiO_{2}}$ slab modes may propagate for somewhat longer distances
before leaking into the underlying silicon substrate of higher refractive
index. Numerical simulations show that the attenuation of slab modes
is $\geq90$($200$)$\,$$\mathrm{dB/cm}$ for a $3$($2$)$\,$$\mu m$
thick $\mathrm{SiO_{2}}$ buffer layer. Hence, detectors can be efficiently
shielded from pump photons propagating inside the cladding layer if
they are separated by a centimeter from the fiber-to-chip interface.

$\mathrm{AlN}$ has a $\chi^{(2)}$ coefficient ($4.7\,\mathrm{pm/V}$)
\cite{Pernice2012_APL} which is almost ten times smaller than that
of $\mathrm{LiNbO_{3}}$ ($41.7\,\mathrm{pm/V}$) \cite{Wang2014_OE}.
However, using advanced nanofabrication techniques we are able to
realize microring resonators with high quality factors, which results
in a resonant pump power enhancement that compensates for the weaker
nonlinear coefficient. This is confirmed by comparing the SHG efficiency
achieved in our $\mathrm{AlN}$-device ($1.16\,\mathrm{W^{-1}}$)
to recently reported SHG efficiencies in $\mathrm{LiNbO_{3}}$ micro
disks ($0.109\,\mathrm{W^{-1}}$) \cite{Wang2014_OE,Lin2015_SR}.
We see potential for further enhancing the nonlinear conversion efficiency
for $\mathrm{AlN}$ microring resonators by decreasing the ring radius
and improving the quality factor of the microring resonator. We believe
that five times higher quality factors and a three-fold decrease in
radius are achievable in future $\mathrm{AlN}$ devices, which brings
SHG efficiencies of $400\,\mathrm{W^{-1}}$ within reach.

\vbox{}

\noindent\textbf{CONCLUSIONS}

\noindent In conclusion, the demonstrated photon pair source based
on the $\chi^{(2)}$ nonlinearity of $\mathrm{AlN}$ microring resonators
and the quantum correlation measurement with waveguide-integrated
single-photon detectors constitute an exciting step towards fully
integrated quantum photonic circuits. Compared to photonic circuits
fabricated from more traditional semiconductor materials, e.g. silicon-on-insulator,
$\mathrm{AlN}$ permits high quality (high brightness, high purity,
low noise) correlated photon pair emission spectrally far separated
from the pump light. $\mathrm{AlN}$-on-insulator therefore holds
great potential for realizing efficient pump suppression and monolithic
integration of non-classical light sources with single photon detectors
on the same chip. Additionally, high-speed phase modulation via the
electro-optic effect will enable real-time circuit reconfiguration.
In combination with high-efficiency single-photon detectors it will
thus be possible to generate non-classical photonic states and implement
feed-forward schemes as well as quantum logic operations in a scalable
manner.

\vbox{}

\noindent\textbf{Acknowledgments}\\ H.X.T. acknowledges support
from a Packard Fellowship in Science and Engineering. C.S. acknowledges
financial support from the Deutsche Forschungsgemeinschaft (SCHU 2871/2-1).
Facilities used for fabrication were supported by Yale SEAS cleanroom
and Yale Institute for Nanoscience and Quantum Engineering. The authors
thank Michael Power and Dr. Michael Rooks for assistance in device
fabrication.

\end{document}